\documentclass[12pt]{article}
\usepackage{amsmath}  
\usepackage{amsthm}
\usepackage{amssymb}
\usepackage{graphicx,psfrag,epsf}
\usepackage{xcolor}
\usepackage{enumerate}
\usepackage{enumitem}
\usepackage{natbib}
\usepackage{hyperref}
\usepackage[ruled,linesnumbered]{algorithm2e}
\usepackage{url} 
\theoremstyle{plain}
\newtheorem{theorem}{Theorem}[section]

\newtheorem{corollary}{Corollary}[section]

\theoremstyle{definition}

\newtheorem{example}{Example}[section]
\newtheorem{assumption}{Assumption}
  \numberwithin{assumption}{subsection}

\theoremstyle{remark}
\newtheorem*{remark}{Remark}

\newcommand{\blind}{1}
\newcommand{\customlabel}[2]{%
   \protected@write \@auxout {}{\string \newlabel {#1}{{#2}{\thepage}{#2}{#1}{}} }%
   \hypertarget{#1}{#2}
}

\begin{document}

\def\spacingset#1{\renewcommand{\baselinestretch}%
{#1}\small\normalsize} \spacingset{1}



\if1\blind
{
  \title{\bf Spatial Confidence Regions for Combinations of Excursion Sets in Image Analysis}
  \author{Thomas Maullin-Sapey\textsuperscript{1}\thanks{\scriptsize{TMS: Thomas.Maullin-Sapey@bdi.ox.ac.uk. AS: armins@ucsd.edu. TEN: Thomas.Nichols@bdi.ox.ac.uk}}, Armin Schwartzman\textsuperscript{2,3} and Thomas E. Nichols\textsuperscript{1}\\ \\
  \small \textsuperscript{1}Big Data Institute, Li Ka Shing Centre for Health Information and Discovery,\\ 
  \small Nuffield Department of
Population Health, University of Oxford, Oxford, UK;\\ 
\small \textsuperscript{2}Division of Biostatistics, University of California, San Diego, CA, USA;\\ 
\small \textsuperscript{3}Halicio\u{g}lu Data Science Institute, University of California, San Diego, USA}
  \maketitle
} \fi

\if0\blind
{
  \bigskip
  \bigskip
  \bigskip
  \begin{center}
    {\LARGE\bf Spatial Confidence Regions for Combinations of Excursion Sets in Image Analysis}
\end{center}
} \fi

\begin{abstract}
\footnotesize{The analysis of excursion sets in imaging data is essential to a wide range of scientific disciplines such as neuroimaging, climatology and cosmology. Despite growing literature, there is little published concerning the comparison of processes that have been sampled across the same spatial region but which reflect different study conditions. Given a set of asymptotically Gaussian random fields, each corresponding to a sample acquired for a different study condition, this work aims to provide confidence statements about the intersection, or union, of the excursion sets across all fields. Such spatial regions are of natural interest as they directly correspond to the questions ``Where do \textit{all} random fields exceed a predetermined threshold?", or ``Where does \textit{at least one} random field exceed a predetermined threshold?". To assess the degree of spatial variability present, we develop a method that provides, with a desired confidence, subsets and supersets of spatial regions defined by logical conjunctions (i.e.~set intersections) or disjunctions (i.e.~set unions), without any assumption on the dependence between the different fields. The method is verified by extensive simulations and demonstrated using a task-fMRI dataset to identify brain regions with activation common to four variants of a working memory task.}
\end{abstract}

\noindent%
{\it Keywords: Spatial Statistics, Excursion Sets, Confidence Regions, Linear Model}  
\vfill

\newpage
\spacingset{1.45} 

\section{Introduction}\label{intro}

The collection and analysis of imaging data, modelled as a random field sampled on a spatial domain, is central to a broad range of scientific disciplines such as neuroimaging, climatology, and cosmology. Often, spatial data drawn from $n$ observations of a random field are combined to obtain an estimate, $\hat{\mu}_n$, of some spatially varying target function $\mu$. The target function $\mu$ is defined on a closed spatial domain, $S\subset\mathbb{R}^N$, and maps spatial locations to some variable of interest in $\mathbb{R}$ (e.g. seismic activity, infrared heat, changes in blood oxygenation in the brain, etc.). In such applications, it is typically desirable to ask ``At which locations does the target function exceed a certain value, $c$?". For example, ``Where has a significant change in temperature occurred?"; ``Where do significant heat readings indicate the presence of celestial bodies?". Such questions are addressed by the study of excursion sets, i.e.~sets of the form $\{s\in S:\mu(s)\geq c\}$.

A wealth of literature has previously focused on documenting the geometric and topological properties of excursion sets of random functions (c.f. \citet{Adler:1981geom}, \citet{Torres1994}, \cite{cao1999}, \citet{Azas2009}, \citet{Adler:2009rand}). These properties include, for example, the Euler characteristic (a measure of topological structure), the Hausdorff dimension (indicating how fractal the set may be), Lipschitz Killing curvatures (describing high-dimensional volumes and areas) and Betti Numbers (describing how many stationary points appear above the threshold, $c$) (\citet{worsley1996}, \citet{Adler1977}, \citet{adler2017estimating}, \citet{Pranav2019}). Much of this work, though, is limited to homogeneous stationary processes, i.e. those with zero mean and a spatial covariance structure which is dependent upon only distance. Only the most recent literature, including the present paper, concerns processes with a non-zero mean function and a potentially heterogeneous covariance function.

In addition, at present there is little published concerning how excursion sets may be compared to one another. In many applications, it is typical for a researcher to collect data from two or more study conditions and contrast the results to draw some meaningful inference (e.g. temperature changes in winter vs summer, heat measurements gathered using different imaging modalities, brain activity in healthy subjects across a range of tasks, etc.). Such settings are akin to having $M$ spatially aligned, potentially correlated, estimates, $\hat{\mu}_n^1,\hat{\mu}_n^2,...\hat{\mu}_n^M:S\rightarrow\mathbb{R}$, of $M$ target functions, $\mu^1,\mu^2,...\mu^M$. 

Often, pertinent questions that arise in such settings may be formally expressed using logical statements which involve conjunctions (logical `and's), negations (logical `not's) and disjunctions (logical `or's). For example, a climatologist may ask where significant changes in temperature were observed during either winter \textit{or} summer (i.e.~``Where does $\mu^1$ \textit{or} $\mu^2$ exceed $c$?'). Alternatively, a neuroscientist may ask ``At which locations in the brain did subjects exhibit signs of cognitive behaviour during one task \textit{and not} another?" (e.g. ``At which locations does $\mu^1$, \textit{and not} $\mu^2$, exceed $c$?").

One approach to answering such questions is to employ null-hypothesis testing. A hypothesis test of a disjunction of nulls to assess evidence for a conjunction of alternative hypotheses can be conducted with an `intersection-union' test.  An intersection-union test rejects at level $\alpha$ when all of the individual nulls are rejected at level $\alpha$. This approach has proven particularly useful for tests of bioequivalence (\citet{Berger1996}, \citet{Berger1997}), but does not consider the spatial aspect that is our focus here. 

It is common practice for researchers to compare images corresponding to different study conditions via rudimentary visual inspection. (In fMRI, just a few recent examples of qualitative assessment of `overlap' and `differences' are \citet{Zhang2021}, \citet{Dijkstra2021}, \citet{Ferreira2021}). Such a subjective practice may lead to biased or misleading results. In recent years, much literature has been published on the theory of confidence regions, which focuses on quantifying spatial uncertainty by providing, with a fixed confidence, sub- and super-sets bounding the excursion set of a single target function (\citet{SSS}, \citet{Bowring2019}, \citet{Bowring2021}). However, the theory of confidence regions does not currently offer a means for investigating logical conjunctions and disjunctions of exceedance statements (i.e.~intersections and unions of excursion sets). This work addresses this issue by first proposing a theory of spatial confidence regions for `conjunction inference' (i.e.~inference for logical statements involving conjunctions) in image analysis. Following this, the proposed method is extended to allow the investigation of statements containing disjunctions and negations.

\begin{figure*}[hbt!]
\center{\includegraphics[width=0.9\textwidth]
{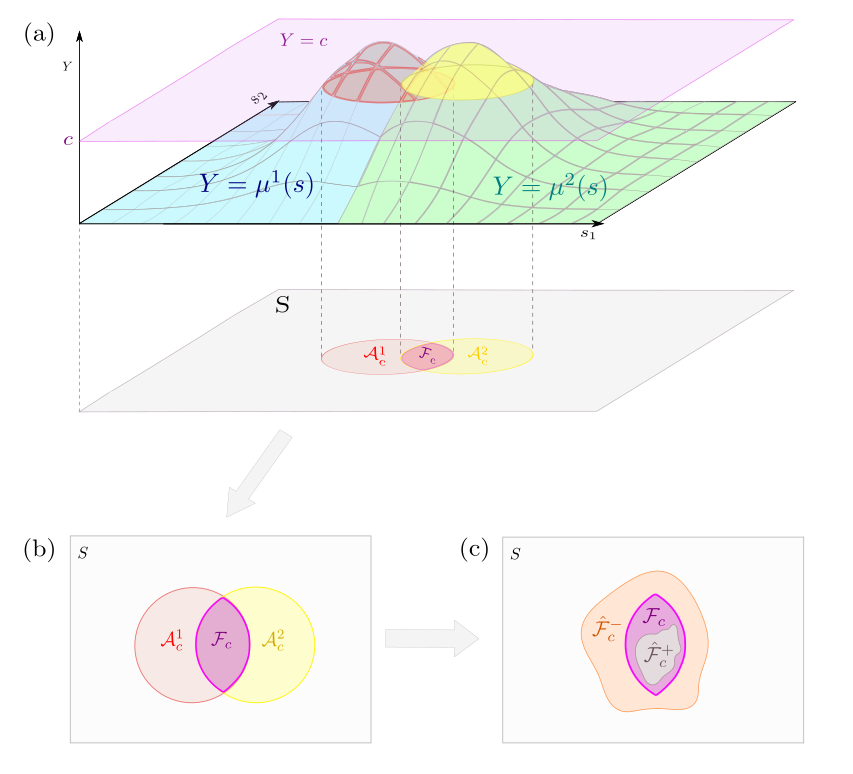}}
\caption{\label{mufig} Illustration of the set $\mathcal{F}_c$ for a setting in which $M=2$. Shown in $(a)$ are two spatially-varying target functions, $\mu^1(s)$ (blue) and $\mu^2(s)$ (green), overlaid and thresholded at the level $c$. In $(b)$, the regions at which $\mu^1(s)\geq c$ and $\mu^2(s)\geq c$ are displayed as red and yellow circles, respectively, with the intersection set, $\mathcal{F}_c$, illustrated in purple. In $(c)$, potential confidence regions, $\hat{\mathcal{F}}_c^{+}$ (grey) and $\hat{\mathcal{F}}_c^{-}$ (orange), for $\mathcal{F}_c$ are illustrated.}
\end{figure*}

Formally, our work primarily focuses on the intersection of $M$ excursion sets (i.e.~the region over which the conjunction statement ``$\mu^1(s)\geq c$ \textit{and} $\mu^2(s)\geq c$ \textit{and} ... \textit{and}  $\mu^M(s)\geq c$" holds). We denote $\mathcal{M}=\{1,...,M\}$, and for each $i\in\mathcal{M}$ (i.e.~for each study condition) define the $i^{th}$ true excursion set and $i^{th}$ estimated excursion set as:
\begin{equation}\nonumber
    \mathcal{A}^i_c=\{s\in S: \mu^i(s)\geq c\}\quad\text{   and   }\quad\hat{\mathcal{A}}^i_c=\{s\in S: \hat{\mu}_n^i(s)\geq c\},
\end{equation}
respectively. Next, we define $\mathcal{F}_c$ to be the intersection of the sets $\{\mathcal{A}_c^i\}_{i \in \mathcal{M}}$, and $\hat{\mathcal{F}}_c$ to be the intersection of the sets $\{\hat{\mathcal{A}}_c^i\}_{i \in \mathcal{M}}$. In other words, the spatial region we are interested in, and our estimate of that region, are given by:
\begin{equation}\nonumber
    \mathcal{F}_c= \bigg\{s\in S: \min_{i\in\mathcal{M}}\mu^i(s)\geq c\bigg\} \quad \text{and}\quad\hat{\mathcal{F}}_c= \bigg\{s\in S: \min_{i\in\mathcal{M}}\hat{\mu}_n^i(s)\geq c\bigg\},
\end{equation}
respectively. An illustration of $\mathcal{F}_c$ in a setting in which $M=2$ is provided by Fig. \ref{mufig}.

We note here that the above construction can be adjusted to allow $c$ to vary across study conditions (i.e.~``$\mu^1(s)\geq c_1$ \textit{and} ... \textit{and}  $\mu^M(s)\geq c_M$" for $c_1,...,c_M\in\mathbb{R}$). Such an adjustment would involve simply translating each field upwards or downwards (e.g.~substituting $\{\mu^i\}_{i\in\mathcal{M}}$ for $\{\mu^i-c+c_i\}_{i\in\mathcal{M}}$). For ease in the proceeding text, we shall assume $c$ is equal for all study conditions.

Our goal is to obtain confidence regions $\hat{\mathcal{F}}^+_c$ and $\hat{\mathcal{F}}^-_c$ such that the below probability holds asymptotically;
\begin{equation}\label{inclusionProb}
    \mathbb{P}\bigg[\hat{\mathcal{F}}^+_c\subseteq \mathcal{F}_c\subseteq\hat{\mathcal{F}}^-_c\bigg]=1 - \alpha
\end{equation} 
for a predefined tolerance level $\alpha$ ($\alpha=0.05$, for example). To generate such regions, we build on the theory of \citet{SSS}, to show that under an appropriate definition of $\hat{\mathcal{F}}^+_c$ and $\hat{\mathcal{F}}^-_c$, the above probability may be approximated using quantiles of a well-defined random variable. Using a wild $t-$bootstrapping procedure, we will then demonstrate that the relationship between this random variable and the above probability can be used to generate $\hat{\mathcal{F}}^+_c$ and $\hat{\mathcal{F}}^-_c$ for any desired value of $\alpha$. Unlike much of the previous literature on random excursion sets, in this work no assumption is made on the processes' means or spatial covariance structure, and we do not make any assumption of between-`study condition' independence (i.e. $\{\hat{\mu}_n^i\}_{i \in\mathcal{M}}$ may be correlated with one another).

In the following sections, we first describe the notation and assumptions upon which our theory relies. Following this, we provide the central theoretical result of this work, which relates Equation (\ref{inclusionProb}) to an exceedance statement for a well-defined random variable. Next, via the use of the wild $t-$bootstrap, we detail how this result may be employed to obtain confidence regions for conjunction inference in the setting of linear regression modelling. Finally, we validate our results with simulations and a real data example based on fMRI data taken from the Human Connectome Project (\citet{Essen2013}). Proofs of the theory presented in this work, alongside further illustration and extensive simulation results, are provided as supplementary material.



\section{Confidence Regions for Excursion Set Combinations}\label{setTheory}

\subsection{Notation}\label{setupsect}

In the following sections, we shall need notation to describe the numerous possible low dimensional sub-manifolds which can arise from intersecting the boundaries of $M$ excursion sets. To do so, we first denote the power set of a finite set, $A$, as $\mathcal{P}(A)$, and define $\mathcal{P}^{+}(A)$ as $\mathcal{P}^{+}(A)=\mathcal{P}(A)\setminus\{\emptyset\}$. For each $i \in \mathcal{M}$, we define $\partial\mathcal{A}_c^i$ as the level set $\{s \in S: \mu^i(s)=c\}$. Similarly, we define $\partial\mathcal{F}_c$ as the level set $\{s \in S: \min_{i\in\mathcal{M}}\mu^i(s)=c\}$. For a spatial set, $B\subseteq S$, its complement in $S$ is denoted $B^c=S\setminus B$, its topological closure is denoted $\overline{B}$, its interior is denoted $B^\circ$ and its boundary is denoted $\partial B$. In addition, for closed $B$, the notation $B^1$ and $B^{-1}$ shall represent $B$ and $\overline{B^c}$, respectively. We note here that the notation $\partial \mathcal{A}^i_c$ and $\partial B$ may conflict with one another. This conflict is resolved, however, by the assumptions of Section \ref{assumpSect} which ensure that the boundaries of $\{\mathcal{A}_c^i\}_{i \in \mathcal{M}}$ and $\mathcal{F}_c$ are equal to the level sets $\{\partial\mathcal{A}_c^i\}_{i \in \mathcal{M}}$ and $\partial\mathcal{F}_c$ (at least locally, in the vicinity of $\partial \mathcal{F}_c$, which is sufficient for our purposes).

\begin{figure*}[hbt!]
\center{\includegraphics[width=0.92\textwidth]
{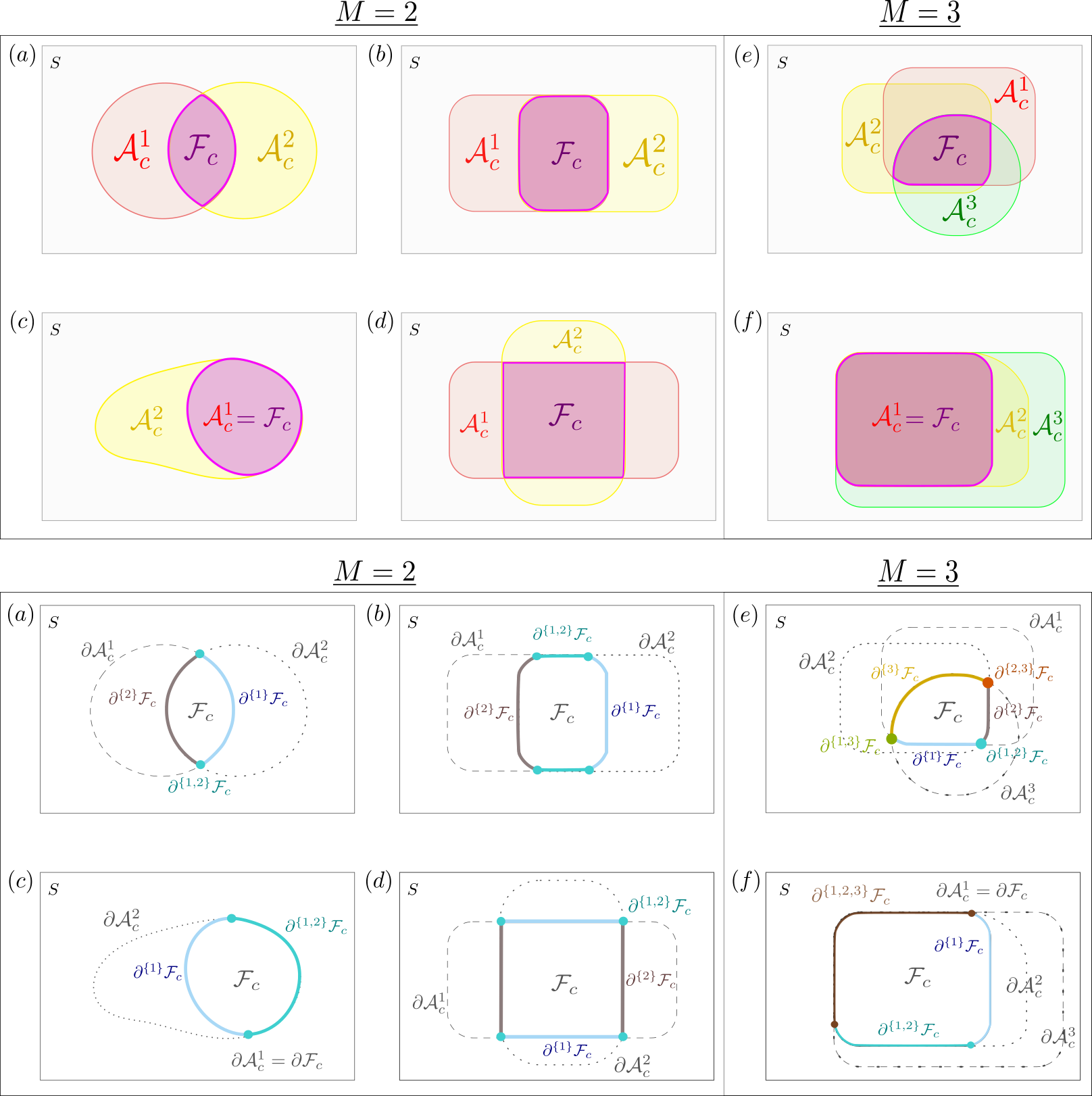}}
\caption{\label{bdryfig} Annotated images of `overlapping' excursion sets (top) with corresponding illustrations of boundary partitions (bottom) for settings in which the number of study conditions, $M$, is equal to $2$ (left) or $3$ (right).}
\end{figure*}

We can now partition the level set $\partial\mathcal{F}_c$ into sub-manifolds using the set of possible intersections of the $M$ excursion set boundaries. For all $\alpha \in \mathcal{P}^+(\mathcal{M})$, we define the boundary segment $\partial^{\alpha}\mathcal{F}_c$ as follows;
\begin{equation}\nonumber
\partial^\alpha \mathcal{F}_c =\partial\mathcal{F}_c \cap \bigg(\bigcap_{i \in \alpha} \partial\mathcal{A}_c^i\bigg) \cap \bigg(\bigcap_{j \in \mathcal{M}\setminus\alpha} (\partial\mathcal{A}_c^j)^c\bigg).
\end{equation}
We note it is possible that, for some $\alpha \in \mathcal{P}^{+}(\mathcal{M})$, the boundary segment $\partial^\alpha \mathcal{F}_c$ is empty (in fact, this is often the case in practice). This notation is crucial to the statement of Theorem \ref{Mainthm} and is illustrated by Fig. \ref{bdryfig}.

\subsection{Assumptions}\label{assumpSect}
In this section, we outline and discuss the assumptions upon which our theory relies. Additional discussion of why each assumption is required, alongside illustration, is given in Supplementary Theory Section S2.

\begin{assumption}\label{clt}
For each $i\in \mathcal{M}$, there exists a bounded function $\sigma^i:S \rightarrow \mathbb{R}^{+}$ and positive sequence $\tau_n\rightarrow 0$, such that the below Central Limit Theorem (CLT) holds:
\begin{equation}\nonumber\bigg\{\frac{\hat{\mu}^i_n(s)-\mu^i(s)}{\tau_n\sigma^i(s)}\bigg\}_{s\in S, i \in \mathcal{M}} \xrightarrow{d} \{G^i(s)\}_{s\in S, i \in \mathcal{M}}
\end{equation}
where $\{G^i(s)\}_{s \in S, i \in \mathcal{M}}$ is a well-defined multi-variate random field with continuous sample paths in $S$ and $\xrightarrow{d}$ represents convergence in distribution.
\end{assumption}

\begin{remark} This assumption introduces the notation $\sigma^i(s)$ and $\tau_n$. In typical applications, $\sigma^i(s)$ represents standard deviation across observations (for study condition $i$, at spatial location $s$) and $\tau_n$ is a decreasing function of sample size. For conciseness in the following text, we now define $\{g^i\}_{i \in \mathcal{M}}$ and $\{\hat{g}_n^i\}_{i \in \mathcal{M}}$ as;
\begin{equation}\nonumber g^i(s)=\frac{\mu^i(s)-c}{\sigma^i(s)}\quad\text{ and }\quad\hat{g}_n^i(s)=\frac{\hat{\mu}_n^i(s)-c}{\sigma^i(s)}.
\end{equation}
\end{remark}

\begin{assumption}\label{ctnty}We assume that: 
\begin{enumerate}[label=(\alph*)]
\setlength\itemsep{0.5em}
\item The functions $\{g^i\}_{i \in \mathcal{M}}$ are continuous on $S$.
\item For $n$ large enough, the functions $\{\hat{g}^i_n\}_{i \in \mathcal{M}}$ are continuous on $S$.
\end{enumerate}
\end{assumption}

\begin{remark} In practice, assumptions of this form are already common in the imaging literature, as in many applications $\{\mu^i\}_{i \in \mathcal{M}}$, $\{\hat{\mu}_n^i\}_{i \in \mathcal{M}}$ and $\{\sigma^i\}_{i \in \mathcal{M}}$ are assumed to be continuous across space. For further remarks, see Supplementary Theory Section S2.2.

\end{remark}

\begin{assumption}\label{diff} We assume that:
\begin{enumerate}[label=(\alph*)]
\setlength\itemsep{0.5em}
\item Every open ball around a point on the boundary $\partial \mathcal{F}_c$ has a non-empty intersection with $(\mathcal{F}_c)^{\circ}$. In addition, for all $\alpha \in \mathcal{P}^{+}(\mathcal{M})$ every open ball around a point in $\partial^{\alpha} \mathcal{F}_c$ has a non-empty intersection with the set:
\begin{equation}\nonumber
\mathcal{J}_c^{\alpha}:=\bigg(\bigcap_{i \in \alpha} (\mathcal{A}^{i}_c)^c\bigg)  \cap \bigg(\bigcap_{j \in \mathcal{M}\setminus\alpha} (\mathcal{A}^{j}_c)^{\circ}\bigg) 
\end{equation}
where, if $\mathcal{M}\setminus\alpha$ is empty, the last term is defined to be equal to $S$.

\item There is an open neighbourhood of $\partial\mathcal{F}_c$ over which the functions $\{g^i\}_{i \in \mathcal{M}}$ and, for $n$ large enough, $\{\hat{g}^i_n\}_{i \in \mathcal{M}}$ are $C^1$ with finite, non-zero, gradients.

\item For every point $s \in \partial\mathcal{F}_c$ and $\alpha\in\mathcal{P}^{+}(\mathcal{M})$, if $s\in\partial\big(\bigcap_{i\in\alpha}\mathcal{A}_c^i\big)$ then the set $\partial\big(\bigcap_{i\in\alpha}\mathcal{A}_c^i\big)$ partitions every sufficiently small open ball around $s$ into exactly two components, each of which is path connected.
\end{enumerate}
\end{assumption}

\begin{remark}
The above statements ensure that $\mathcal{F}_c$ is non-empty and has a well defined boundary which is equal to the level set $\partial\mathcal{F}_c=\{s\in S: \min_{i \in\mathcal{M}}g^i(s)=0\}$. All three statements ensure that no changes in topology occur at the level $c$ by requiring that $\partial \mathcal{F}_c$ does not contain plateaus (spatial regions of non-zero measure over which $\min_{i\in\mathcal{M}}g^i(s)=0$ exactly), local minima or maxima. In addition, each statement handles pathological cases which the other statements do not. For brevity, we do not list such cases here but instead provide further discussion in Supplementary Theory Section S2.3.

\end{remark}

Each of the examples presented in Fig. \ref{bdryfig} satisfy Assumptions \ref{ctnty} and \ref{diff} and highlight several features worthy of note. Firstly, we do not assume $\partial \mathcal{F}_c$ is a $C^1$ curve, but rather piece-wise $C^1$ (c.f. examples $(d)$ and $(e)$). Secondly, the assumptions allow for the possibility that the excursion sets could be nested within one another (c.f. examples $(c)$ and $(f)$). Thirdly, the assumptions permit the excursion sets to share common boundaries (see examples $(b)$, $(c)$ and $(f)$). The last two observations are of practical relevance as, in many applications, it may be expected that the target functions for different study conditions exhibit a strong similarity. For instance, in the neuroimaging example, it may be expected that some anatomical regions of the brain display evidence of activation for all of the study conditions. We note here that there is no requirement that $\mathcal{F}_c$ be (globally) path-connected; $\mathcal{F}_c$ may consist of several disconnected components without violating the assumptions of this section.

\subsection{Theory}\label{theory}

In this section, we present the central result of this work, Theorem \ref{Mainthm}, alongside two corollaries, Corollary \ref{corr1} and Corollary \ref{corr2}. To do so, we now define the nested sets $\hat{\mathcal{F}}^{-}_c$ and $\hat{\mathcal{F}}^{+}_c$ by thresholding the statistic $\tau_n^{-1}\min_{i\in\mathcal{M}}\hat{g}^i_n(s)$:
\begin{equation}\nonumber
\hat{\mathcal{F}}^{-}_c:=\hat{\mathcal{F}}^{-}_c(a)=\{s \in S : \tau_n^{-1}\min_{i\in\mathcal{M}}\hat{g}^i_n(s) \geq -a\},
\end{equation}
\begin{equation}\nonumber
\hat{\mathcal{F}}^{+}_c:=\hat{\mathcal{F}}^{+}_c(a)=\{s \in S : \tau_n^{-1}\min_{i\in\mathcal{M}}\hat{g}^i_n(s) \geq +a\},
\end{equation}
using some constant $a\in\mathbb{R}^{+}$. To find an appropriate value of $a$, such that Equation (\ref{inclusionProb}) holds for a desired tolerance level (e.g. $\alpha=0.05$), Theorem \ref{Mainthm} is required.


\begin{theorem}\label{Mainthm}
Under the assumptions of Section \ref{assumpSect}, the below holds:
\begin{equation}\label{Mainthmstatement}\lim_{n \rightarrow \infty}\mathbb{P}[\hat{\mathcal{F}}^{+}_c\subseteq\mathcal{F}_c\subseteq\hat{\mathcal{F}}^{-}_c]=\mathbb{P}[H\leq a],\end{equation}
where the variable $H$ is defined as follows;
\begin{equation}\nonumber
H=\max_{\alpha \in \mathcal{P}^+(\mathcal{M})}\bigg(\sup_{s \in \partial^\alpha \mathcal{F}_c} \big|\min_{i \in \alpha}(G^i(s))\big|\bigg).
\end{equation}
\end{theorem}
Conceptually, the event $\{H> a\}$ may be thought of as the situation in which, at some point, $s$, inside some boundary segment, $\partial^\alpha \mathcal{F}_c$, a large value was observed for the statistic $|\min_{i \in \alpha}(G^i(s))|$. Therefore, Theorem \ref{Mainthm} tells us the probability that the statement $\hat{\mathcal{F}}^{+}_c\subseteq\mathcal{F}_c\subseteq\hat{\mathcal{F}}^{-}_c$ is violated is directly determined by such points. If we can find a value of $a$ such that $\mathbb{P}[H \leq a]=1-\alpha$ then asymptotically, the inclusion probability, $\mathbb{P}[\hat{\mathcal{F}}^+_c\subseteq \mathcal{F}_c\subseteq\hat{\mathcal{F}}^-_c]$, will also equal $1-\alpha$. In other words, if the $(1-\alpha)^{th}$ quantile of the distribution of $H$ is known, then confidence regions $\hat{\mathcal{F}}^-_c$ and $\hat{\mathcal{F}}^+_c$ may be constructed which satisfy Equation (\ref{inclusionProb}). We shall take up the task of evaluating the quantiles of $H$ in Section \ref{LM}.

Theorem \ref{Mainthm} is key to generating confidence regions for conjunction inference (intersections of excursion sets). However, Theorem \ref{Mainthm} may also be extended to allow investigation of other logical statements involving negations or disjunctions. Below we provide two corollaries, each of which is illustrated via a worked example. Corollary \ref{corr1} extends Theorem \ref{Mainthm} to account for logical negations, whilst Corollary \ref{corr2} provides an equivalent statement for inference performed upon logical disjunctions (unions of excursion sets). 

\begin{example} Suppose, given two target functions, $\mu^1$ and $\mu^2$, we were interested in the region defined by the logical conjunction `$\mu^1\geq c$ \textit{and} $\mu^2\leq c$' (i.e.~``Where did a variable of interest exceed a threshold under one study condition \textit{and not} under another?"). In the notation of Section \ref{setupsect}, this region is readily seen to be given by $(\mathcal{A}_c^1)^1 \cap (\mathcal{A}_c^2)^{-1}$. 

To apply the result of Theorem \ref{Mainthm}, we first define $\tilde{\mu}^1=\mu^1$ and $\tilde{\mu}^2=2c-\mu^2$ and note that the logical statement `$\mu^1\geq c$ \textit{and} $\mu^2\leq c$' is identical to the statement `$\tilde{\mu}^1\geq c$ \textit{and} $\tilde{\mu}^2\geq c$'. By using the latter statement, Theorem \ref{Mainthm} may now be applied to obtain the desired confidence regions (assuming that an appropriate mechanism exists for evaluating the quantiles of $H$). However, during this process, the limiting variables $\{G^i\}_{i \in\{1,2\}}$, which correspond to the target functions $\{\mu^i\}_{i \in\{1,2\}}$, must be replaced by variables corresponding to the functions $\{\tilde{\mu}^i\}_{i \in\{1,2\}}$, $\{\tilde{G}^i\}_{i \in\{1,2\}}$. By noting the definitions of $\{\tilde{\mu}^i\}_{i \in\{1,2\}}$ and $\{\tilde{G}^i\}_{i \in\{1,2\}}$, it can be seen that $\tilde{G}^1=G^1$ and  $\tilde{G}^2=-G^2$. 

In summary, a procedure for generating confidence regions corresponding to the logical conjunction `$\mu^1\geq c$ \textit{and} $\mu^2\leq c$' would be identical to that previously described, except that $G^2$ would be substituted for $-G^2$ and $\partial \mathcal{F}_c$ would be substituted for $\partial ((\mathcal{A}^1_c)^1\cap(\mathcal{A}^2_c)^{-1})$. In general, this example may be extended to allow for inference to be performed upon arbitrary conjunctions of statements and negated statements. To achieve this, the definitions of $\mathcal{F}_c$, $\hat{\mathcal{F}}^{-}_c$ and $\hat{\mathcal{F}}^{+}_c$ must be extended as follows.
\end{example}

\begin{corollary}\label{corr1} Let $\{\delta_i\}_{i \in \mathcal{M}}$ be an arbitrary sequence of integers in $\{-1,1\}$ and define $\mathcal{F}_c$ as $\mathcal{F}_c=\bigcap_{i \in \mathcal{M}} (\mathcal{A}_c^i)^{\delta_i}$. Similarly, extend the definition of the sets $\hat{\mathcal{F}}^{+}_c$ and $\hat{\mathcal{F}}^{-}_c$ to:
\begin{equation}\nonumber
    \hat{\mathcal{F}}^{+}_c=\bigcap_{i \in \mathcal{M}}\{s \in S :\tau_n^{-1}\hat{g}^i_n(s)\geq +a\}^{\delta_i} \quad \text{and}\quad\hat{\mathcal{F}}^{-}_c=\bigcap_{i \in \mathcal{M}}\{s \in S :\tau_n^{-1}\hat{g}^i_n(s)\geq -a\}^{\delta_i},
\end{equation}
respectively and, for each $\alpha \in \mathcal{P}^{+}(\mathcal{M})$, define $\partial^\alpha\mathcal{F}_c$ as before.
If $\mathcal{F}_c$ satisfies the assumptions of Section \ref{assumpSect}, then the below statement holds asymptotically:
\begin{equation}\nonumber\lim_{n \rightarrow \infty}\mathbb{P}[\hat{\mathcal{F}}^{+}_c\subseteq\mathcal{F}_c\subseteq\hat{\mathcal{F}}^{-}_c]=\mathbb{P}\bigg[\max_{\alpha \in \mathcal{P}^+(\mathcal{M})}\bigg(\sup_{s \in \partial^\alpha \mathcal{F}_c} \big|\min_{i \in \alpha}(\delta_iG^i(s))\big|\bigg)\leq a\bigg]\end{equation}

\end{corollary}

\begin{example} 
Suppose, given two target functions, $\mu^1$ and $\mu^2$, interest lay in the logical disjunction of `$\mu^1 \geq c$ \textit{or} $\mu^2 \geq c$' (i.e.~``Where did \textit{at least one} of the variables of interest exceed the threshold?"). The region of space over which this statement holds is given by $\mathcal{A}_c^1 \cup \mathcal{A}_c^2$, and shall be denoted $\mathcal{G}_c$.

To generate confidence regions for $\mathcal{G}_c$, we first assume that $\mathcal{G}_c$ satisfies Assumptions \ref{clt}-\ref{diff}. By Assumptions \ref{ctnty} and \ref{diff} and De Morgan's law, it can be seen that $\mathcal{G}_c=((\mathcal{A}_c^1)^{-1} \cap (\mathcal{A}_c^2)^{-1})^{-1}$. It must be noted that Assumptions  \ref{ctnty} and \ref{diff} are necessary for this statement to hold due to the fact that $(\mathcal{A}_c^i)^{-1}=\overline{(\mathcal{A}_c^i)^c}$ and not $(\mathcal{A}_c^i)^c$. By employing Corollary \ref{corr1}, for a fixed value of $\alpha$, confidence regions, $\hat{\mathcal{F}}^{-}_c$ and $\hat{\mathcal{F}}^{+}_c$, may be obtained which satisfy the following;
\begin{equation}\nonumber \lim_{n \rightarrow\infty}\mathbb{P}[(\hat{\mathcal{F}}^{-}_c)^c\subseteq ((\mathcal{A}_c^1)^{-1} \cap (\mathcal{A}_c^2)^{-1})^c \subseteq(\hat{\mathcal{F}}^{+}_c)^c]=1-\alpha.\end{equation}

By taking the closure of each of the sets in the above probability, and again via careful consideration of Assumptions \ref{ctnty} and \ref{diff}, the below is obtained;
\begin{equation}\nonumber \lim_{n\rightarrow\infty}\mathbb{P}[(\hat{\mathcal{F}}^{-}_c)^{-1}\subseteq \mathcal{G}_c \subseteq(\hat{\mathcal{F}}^{+}_c)^{-1}]=1-\alpha.\end{equation}
In other words, the sets $(\hat{\mathcal{F}}^{-}_c)^{-1}$ and $(\hat{\mathcal{F}}^{+}_c)^{-1}$ serve as confidence regions for disjunction inference on $\mu^1$ and $\mu^2$. Note that, as in the previous example, great attention must paid to the signs which appear in front of the variables $\{G^i\}_{i \in\mathcal{M}}$ in the definition of $H$. As this example began by generating confidence regions for $(\mathcal{A}_c^1)^{-1} \cap (\mathcal{A}_c^2)^{-1}$, it can be seen that the variables $G^1$ and $G^2$ must be appended with negative signs (i.e.~when applying the result of Corollary \ref{corr1}, both $\delta_1$ and $\delta_2$ were set to $-1$). This example may be expanded upon to obtain similar results for larger values of $M$, as shown by Corollary \ref{corr2}. 
\end{example}

\begin{corollary}\label{corr2} Let $\{\delta_i\}_{i \in \mathcal{M}}$ be an arbitrary sequence of integers in $\{-1,1\}$ and define $\mathcal{G}_c$ as $\mathcal{G}_c=\bigcup_{i \in \mathcal{M}} (\mathcal{A}_c^i)^{\delta_i}$. Define the sets $\hat{\mathcal{G}}^{+}_c$ and $\hat{\mathcal{G}}^{-}_c$ as:
\begin{equation}\nonumber
    \hat{\mathcal{G}}^{+}_c=\bigg(\bigcap_{i \in \mathcal{M}}\{s \in S : \tau_n^{-1}\hat{g}^i_n(s)\geq -a\}^{\delta_i}\bigg)^{-1} \quad \text{and}\quad\hat{\mathcal{G}}^{-}_c=\bigg(\bigcap_{i \in \mathcal{M}}\{s \in S : \tau_n^{-1}\hat{g}^i_n(s)\geq +a\}^{\delta_i}\bigg)^{-1},
\end{equation}
respectively and, for each $\alpha \in \mathcal{P}^{+}(\mathcal{M})$, define $\partial^\alpha\mathcal{G}_c$ analogously to $\partial^\alpha\mathcal{F}_c$ in the previous sections. If $\mathcal{G}_c$ satisfies the assumptions of Section \ref{assumpSect}, then the below statement holds asymptotically:
\begin{equation}\nonumber\lim_{n \rightarrow \infty}\mathbb{P}[\hat{\mathcal{G}}^{+}_c\subseteq\mathcal{G}_c\subseteq\hat{\mathcal{G}}^{-}_c]=\mathbb{P}\bigg[\max_{\alpha \in \mathcal{P}^+(\mathcal{M})}\bigg(\sup_{s \in \partial^\alpha \mathcal{G}_c} \big|\min_{i \in \alpha}(-\delta_iG^i(s))\big|\bigg)\leq a\bigg].\end{equation}
\end{corollary}

\section{Spatial Conjunction Inference for the Linear Model}\label{LM}

In this section, we build upon the work of \citet{SSS} and \citet{Bowring2019}, in which methods for generating confidence regions were presented for a single target function, $\mu(s)$, derived from a linear regression model. In Section \ref{modSpec} we formally describe the spatially-varying Linear Model (LM) for $M$ study conditions and, in Section \ref{boot}, we explain how the wild $t$-bootstrap may be applied to estimate quantiles of the variable $H$. Further implementation details, for when data are sampled from a discrete lattice rather than across a continuous space, alongside pseudocode, are provided in Supplementary Results Section S$2$. This section focuses solely on conjunction inference. However, the methods described here may equally be applied to perform inference on other logical statements via the use of the corollaries and examples presented in Section \ref{theory}. 

\subsection{Model Specification}\label{modSpec}

For each $i\in\mathcal{M}$ (i.e.~for each study condition), we assume that the data follow a spatially-varying Linear Model (LM) defined at location $s \in S$ as:
\begin{equation}\label{lmspec}
    Y^i(s) = X^i\beta^i(s) + \epsilon^i(s), \quad \epsilon^i(s) \sim N(0,\Sigma^i(s)).
\end{equation}
The known quantities in the model are; the $(n \times 1)$ vector of responses, $Y^i(s)$, and the $(n \times p)$ design matrix, $X^i$. The unknown model parameters are; the $(p \times 1)$ vector of regression coefficients, $\beta^i(s)$, and the $(n\times n)$ random error covariance matrix, $\Sigma^i(s)$. For each spatial location, $s\in S$, the Generalized Least Squares (GLS) estimator of the parameter vector $\beta^i(s)$, $\hat{\beta}^i(s)$, is given by:
\begin{equation}\nonumber
    \hat{\beta}^i(s) = (X^{i'}\Sigma^i(s)^{-1}X^i)^{-1}X^{i'}\Sigma^i(s)^{-1}Y^i(s).
\end{equation}
In this context, under the $i^{th}$ study condition, at each spatial location $s$, our interest lies in assessing whether linear relationships hold between the elements of the parameter vector $\beta^i(s)$. Such relationships may be expressed using a contrast vector $L^i$ of dimension $(p \times 1)$. In the notation of the previous sections, the target and estimator functions for the spatially-varying LM are given as $\mu^i(s)=L^{i'}\beta^i(s)$ and $\hat{\mu}^i_n(s)=L^{i'}\hat{\beta}^i(s)$ respectively, and $\tau_n^{-1}\sigma^i(s)$ is the contrast standard error, given as: 
\begin{equation}\nonumber\tau_n^{-1}\sigma^i(s)=\big[L^{i'}(X^{i'}\Sigma^i(s)^{-1}X^i)^{-1}L^i\big]^{\frac{1}{2}}.
\end{equation}

Whilst in the above notation we have presented $\{\hat{\beta}^i(s)\}_{i \in \mathcal{M}}$ as being estimated separately using $M$ different models, we note that nothing in our theory prevents the same model being used across all $M$ study conditions. For example, it is possible for the model matrices, $\{Y^i(s)\}_{i \in \mathcal{M}}$ and $\{X^i\}_{i \in \mathcal{M}}$, and parameter vector, $\{\beta^i(s)\}_{i \in \mathcal{M}}$, to be equal for all $i \in \mathcal{M}$, with the only distinction between study conditions being represented by the contrast vector $L^i$. This observation is noteworthy as, in many applications, it is common for researchers to compile all study conditions into a single LM to obtain a heightened statistical power.


To apply Theorem \ref{Mainthm} to the above LM, we now assume that Assumptions \ref{clt}-\ref{diff} hold. Such assumptions are commonly satisfied by standard conditions placed on the continuity and boundedness of the increments and moments of $\{\beta^i(s)\}_{s \in S}$ and $\{\epsilon^i(s)\}_{s \in S}$ (see \citet{SSS} for further detail). We note here that the covariance matrix, $\Sigma^i(s)$, is usually unknown in practice, meaning the function $\tau_n^{-1}\sigma^i(s)$ cannot be calculated. As demonstrated in \citet{SSS}, however, $\Sigma^i(s)$ can be replaced by any consistent estimator $\hat{\Sigma}^i(s)$ and the assumptions given in Section \ref{assumpSect} are still satisfied. Such estimation is common in the statistics literature and is typically achieved by assuming that some fixed correlation structure applies to $\Sigma^i(s)$ (e.g. diagonal independence, auto-regressive, etc.) so that the number of independent variance parameters which must be estimated is small.


\subsection{The Wild $t-$bootstrap}\label{boot}

In practice, the distribution of $H$ is unknown and must be estimated. In this section, for the model specification described in Section \ref{modSpec}, we employ a wild $t-$bootstrap resampling scheme to obtain the quantile, $a$, of $H$ which satisfies $\mathbb{P}[H\leq a]=1-\alpha$.

To do so, we first define the $(n \times 1)$-dimensional residual vector, $R^i$, for the LM of the $i^{th}$ study condition ($i \in \mathcal{M}$) as:
\begin{equation}\nonumber
[R^i_1(s),..., R^i_n(s)]' = R^i(s) = [\hat{\Sigma}^i(s)]^{-\frac{1}{2}}(Y^i(s)-X^i\hat{\beta}^i(s))
\end{equation}
The wild $t-$bootstrap proceeds by, in each bootstrap instance, generating $n$ i.i.d. Rademacher random variables, $\{r_1,...,r_n\}$, (random variables which take the values $-1$ and $+1$ with equal probability) independently of the data. For the $i^{th}$ study condition, at spatial location $s$, a new bootstrap sample is then obtained by multiplying the elements of the residual vector by the Rademacher variables (i.e.~the new sample is given by $\{r_1R^i_1(s),...,r_nR^i_n(s)\}$). Once the boostrap sample has been generated, the standard deviation of the bootstrap sample, $\hat{\sigma}^{*,i}(s)$, is calculated. A bootstrap instance of the variable $G^i$, $\tilde{G}^i$, is now generated as follows;
\begin{equation}\nonumber
\tilde{G}^i(s)=n^{-\frac{1}{2}}\sum_{l=1}^n \frac{r_lR^i_l(s)}{\hat{\sigma}^{*,i}(s)}.
\end{equation}
A bootstrap instance of the variable $H$, $\tilde{H}$, may then be computed using the bootstrap variables $\{\tilde{G}^i(s)\}_{s \in S, i \in \mathcal{M}}$ as follows:
\begin{equation}\label{estH}
    \tilde{H}=\max_{\alpha \in \mathcal{P}^+(\mathcal{M})}\bigg(\sup_{s \in \partial^\alpha \hat{\mathcal{F}}_c} \big|\min_{i \in \alpha}(\tilde{G}^i(s))\big|\bigg)
\end{equation}
Quantiles of the distribution of $H$ may now be estimated empirically from the observed sampling distribution of $\tilde{H}$. Discussion of how the above supremum, taken across continuous space, is evaluated in practice is deferred to Supplementary Results Section S$2.1$. However, we do briefly note here that, as the location of the true boundary segment $\partial^\alpha \mathcal{F}_c$ is in practice unknown, the boundary segment $\partial^\alpha \mathcal{F}_c$ has been replaced by an estimate $\partial^\alpha \hat{\mathcal{F}}_c$. It should be noted that, as a result of this substitution, our proposed method may not be applied when $\hat{\mathcal{F}}_c$ is empty.

Several strong references exist that argue and verify the correctness of using the wild $t-$bootstrap for estimating quantiles of maxima distributions in the above manner. Of particular note are the works of \citet{CHANG2017} and \citet{Bowring2019}, in which the wild $t-$bootstrap is proposed for the estimation of Gaussian maxima distributions and the generation of confidence regions, respectively. Discussion of the wild $t-$bootstrap in an imaging context may also be found in \citet{Telschow2021}. Other significant references which serve as precursors to this work include \citet{Chernozhukov2013} and \citet{SSS}, in which the Gaussian multiplier bootstrap was investigated for estimating Gaussian maxima distributions and generating confidence regions, respectively. More general introductions to such bootstrapping procedures may be found in \citet{Wu1986}, \citet{Efron1994} and \citet{hesterberg2014teachers}.

In the form that it has been presented here, the wild $t-$bootstrap requires that $\{G^i\}_{i \in \mathcal{M}}$ be symmetric. Further, as noted above, the wild $t-$bootstrap has been extensively verified for estimating maxima distributions only in settings in which $\{G^i\}_{i \in \mathcal{M}}$ are Gaussian. Here we stress that, whilst the theory presented in Section \ref{setTheory} makes no assumptions on the distribution of $\{G^i\}_{i \in \mathcal{M}}$, the bootstrap theory presented throughout Section \ref{LM} requires $\{G^i\}_{i \in \mathcal{M}}$ to be Gaussian. In order to utilise the results of Section \ref{setTheory} when the random variables $\{G^i\}_{i \in \mathcal{M}}$ are not Gaussian, alternative methods must be employed for estimating the quantiles of $H$.


\section{Simulations}\label{sim}
\subsection{Simulation Settings}\label{simMeth}

To assess the correctness and performance of the method, a range of simulations were conducted using synthetic data. Each simulation was designed to investigate how the empirical coverage (i.e.~the proportion of simulation instances in which the inclusion $\hat{\mathcal{F}}^{+}_c\subseteq\mathcal{F}_c\subseteq\hat{\mathcal{F}}^{-}_c$ held) was affected by various secondary factors of practical interest. In this section, we describe three simulations designed to investigate how the methods' performance was influenced by; $(1)$ the degree of overlap between excursion sets, $(2)$ the presence of correlation between the noise fields for different study conditions, and $(3)$ the magnitude of the spatial rate of change of the signal. 

In all simulations, for $i \in \mathcal{M}$, the model used to generate the synthetic data took the form;
\begin{equation}\label{simpleLM}
    Y^i(s) = \mu^i(s)+\epsilon^i(s), \quad \epsilon^i(s) \sim N(0,\sigma^i(s)I_n).
\end{equation}
where $n$ was allowed to vary from $40$ to $500$ (in increments of $20$). The aim, across all simulations, was to assess the performance of the method when employed for conjunction inference on the true signal $\{\mu^i\}_{i \in \mathcal{M}}$ (i.e.~when asked the question ``Where do \textit{all} of the $\{\mu^i\}_{i \in \mathcal{M}}$ exceed a predefined threshold?'').


The mechanism used to generate the true signal, $\{\mu^i\}_{i \in \mathcal{M}}$, varied across simulations. In Simulations 1 and 2, $\{\mu^i\}_{i \in \mathcal{M}}$ were generated using two binary images of circles and squares, respectively, positioned in a `Venn diagram' arrangement. Each binary image was scaled by a predefined amount and then smoothed using an isotropic Gaussian filter with a Full-Width Half Maximum (FWHM) of $5$ pixels. In Simulation $3$, $\mu^1$ and $\mu^2$ were simulated as a horizontal and vertical linear ramp, respectively, with predefined gradients.

Each simulation was performed twice, once using noisier `low-Signal-to-Noise Ratio (SNR)' synthetic data and again using less noisy `high-SNR' synthetic data.  To generate the high-SNR synthetic data for Simulations $1$ and $2$, all simulated $\{\mu^i\}_{i \in \mathcal{M}}$ were scaled by a factor of $3$ prior to smoothing. This data generation process is identical to that employed in \citet{Bowring2019}, in which it is noted that synthetic data generated using these parameters strongly resembles that of an fMRI analysis when voxels are of dimension $2$mm$^3$. In Simulation $3$, for the high-SNR data, the linear ramps were initially simulated with a gradient of $8$ per $50$ pixels. Across all simulations, the low-SNR data was generated by reducing the signal magnitude of the high-SNR data by a factor of $4$. In all simulations, thresholds of $c=1/2$ and $c=2$ were employed for the low-SNR data and high-SNR data, respectively. 

As Simulation $1$ aimed to investigate the method's performance as the overlap between excursion sets increased, in this simulation, the distance between the centre of the circles was varied, ranging from $0$ pixels (full overlap) to $50$ pixels (barely overlapping) in increments of $2$ pixels. Similarly, as Simulation $2$ aimed to assess the effect of correlation between $\epsilon^1$ and $\epsilon^2$, in this simulation, this correlation was varied between $-1$ and $1$ in increments of $0.1$ (in all other simulations, the noise fields were generated independently). As Simulation $3$ served to assess how sensitive the empirical coverage was to the rate of change of $\mu^1$ and $\mu^2$ over space, in this simulation, the initial ramp gradients were multiplied by a factor of $k$, where $k$ was varied from $0.25$ to $1.75$ in $0.05$ increments.

All simulation results were obtained as averages taken across $2500$ simulation instances and compared to the nominal coverage using $95\%$ binomial confidence intervals. In all simulation instances, the image dimensions were $(100 \times 100)$ pixels, confidence regions were computed for a tolerance level of $\alpha=0.05$ using $5000$ bootstrap realizations, $\sigma^i(s)$ was computed using the ordinary least squares estimator and $\tau_n$ was computed as $\tau_n=n^{-0.5}$. In each simulation instance, the assessment of whether the inclusion statement, $\hat{\mathcal{F}}^{+}_c\subseteq\mathcal{F}_c\subseteq\hat{\mathcal{F}}^{-}_c$, held was verified using the interpolation-based methods of \citet{Bowring2019}. In this approach, the inclusion statement was deemed to hold if, and only if, the sets of pixels which were identified as $\hat{\mathcal{F}}_c^{-}$, $\mathcal{F}_c$ and $\hat{\mathcal{F}}_c^{+}$ were appropriately nested and, in addition, interpolation along the boundary $\partial\mathcal{F}_c$ confirmed that no violations of the inclusion statement had occurred.

A further six simulations which investigated the influence of other secondary factors of interest were also conducted. The secondary factors considered by these simulations include; the presence of spatial structure in the noise, the effect of $\partial^{\{1\}}\mathcal{F}_c$ and $\partial^{\{2\}}\mathcal{F}_c$ having differing lengths, the effect of $\epsilon^1$ having a much larger variance than $\epsilon^2$ and the impact of varying the value of $M$. A full discussion of these simulations is deferred to Supplementary Results Section S$4$. In addition, in keeping with the previous literature, tolerance levels of $\alpha=0.1$ and $\alpha=0.2$, as well as the substitution of $\partial\mathcal{F}_c$ for $\partial\hat{\mathcal{F}}_c$ in Equation (\ref{estH}), were also considered for simulation (c.f. \citet{SSS}, \citet{Bowring2019}, \citet{Bowring2021}). Again, the results of these simulations may be found in the Supplementary Results document in Sections S$5$ and S$6$. 

\subsection{Simulation Results}\label{simRes}

For a majority of the simulations conducted, the empirical coverage estimates were tightly distributed around the expected nominal coverage. In particular, for the high-SNR data, across all simulations, the $95\%$ binomial confidence intervals consistently captured the nominal coverage at the predicted rate. However, when the simulations employed the low-SNR data, it can be seen that the confidence regions produced by the method were conservative when the data were generated under certain unfavourable conditions.

\begin{figure*}[hbt!]
\center{\includegraphics[width=0.48\textwidth]
{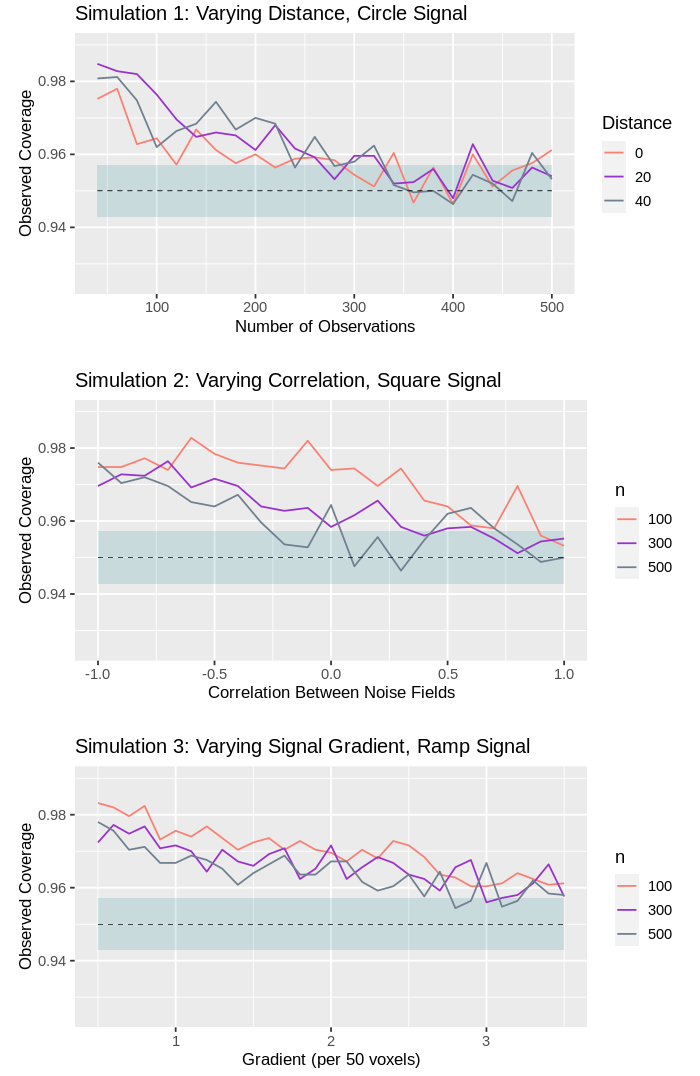}}
\caption{\label{sim1} Simulation results generated using the low-SNR synthetic data.  Nominal coverage is shown as a dashed line, with a corresponding binomial confidence interval also shaded. All data points are averages taken across $2500$ simulation instances.}
\end{figure*}
For example, in Fig. \ref{sim1}, it can be seen that some over-coverage was observed for Simulation $1$ when the data was noisy, and the number of subjects was low. This observation is corroborated by the results of the remaining low-SNR synthetic data simulations (c.f. Supplementary Results Section S$5$). However, in all cases, the degree of agreement between the empirical and nominal coverage improved as the sample size increased. Furthermore, no such over-coverage was observed for the equivalent high-SNR simulations. This observation matches expectation, as Theorem \ref{Mainthm} holds only asymptotically, and is supported by the previous literature on confidence regions, in which similar results were observed for the single-`study condition' setting (c.f. \citet{SSS}, \citet{Bowring2019}).

The presence of strong negative correlation between the study conditions and the rate at which the signal varied across space both also appeared to influence the methods' performance (c.f. Fig \ref{sim1}, Simulations 2 and 3). In both instances, it is likely that the observed over-coverage is due to difficulties in estimating the true location of the boundary $\partial \mathcal{F}_c$, as each of these factors make it harder to identify the locations at which small changes in $\min_{i \in \mathcal{M}}\mu^i$ occur. When high-SNR data was used in the place of low-SNR data, both of these factors had a substantially reduced impact on the empirical coverage.

Despite the above observations, the simulation results overwhelmingly supported the claim that the proposed method is robust to a range of secondary factors. Included in the list of such factors are; variation in the shape and size of $\mathcal{F}_c$, spatial correlation in the noise, variation in the lengths of individual boundary segments, between-`study condition' correlation, realistic levels of variation in the magnitude of the noise, large numbers of study conditions and situations in which boundary segments are shared by many excursion sets. For further detail, see Supplementary Results Sections S$5$ and S$6$.

In terms of time efficiency, the wild $t-$bootstrap provided extremely fast computational performance. For instance, using an Intel(R) Xeon(R) Gold 6126 2.60GHz processor with 16GB RAM, and averaging over the $2500$ simulation instances conducted for $n=500$ high-SNR observations, in Simulation $1$, the time taken to generate confidence regions for a circle separation of $20$ pixels was $5.98$ seconds. For a comprehensive summary of computation times, see Supplementary Results Section S$7$.

\section{Real Data Application}\label{RealDat}

To demonstrate the method in practice, we apply it to fMRI data drawn from the Human Connectome Project (HCP) dataset (\citet{Essen2013}). In this example, task fMRI data were collected as part of a block design from $80$ healthy, unrelated, young adults as part of a working memory task that had four distinct components, each using a different stimuli type: pictures of places, tools, faces and body parts. In Supplementary Results Section S$3$, we provide a brief overview of the imaging acquisition protocol, task paradigm, preprocessing stages and first-level analysis employed for generating this dataset (for exhaustive detail, see \citet{BARCH2013} and \citet{GLASSER2013}). Following first-level analysis, the data consisted of $4$ images for each of the $80$ subjects, measuring the \%BOLD response to each of the $4$ stimuli types. In this example, our interest lies in identifying, at the group-level, which regions of the brain are associated with working memory regardless of stimuli type (i.e.~``Which regions are active in response to \textit{all} four working memory stimuli types?'').

As this work has predominantly focused on two-dimensional excursion sets, a single slice of the brain was chosen for analysis ($z=46$mm, covering portions of the frontal gyrus involved in working memory). For each stimuli type, an $n=80$ group-level linear model was constructed.
Using the proposed method, confidence regions were then generated at the $5\%$ confidence level to assess where the group-level percentage BOLD change exceeded $1\%$ for \textit{all} four stimuli types. To compare the proposed method to standard fMRI inference procedures, a group-level contrast was also generated using the Big Linear Model toolbox for each of the four stimuli types. In addition, single-`study condition' confidence regions were also generated using the methods of \citet{SSS} for each of the four stimuli types. 

\begin{figure*}[hbt!]
\center{\includegraphics[width=\textwidth]
{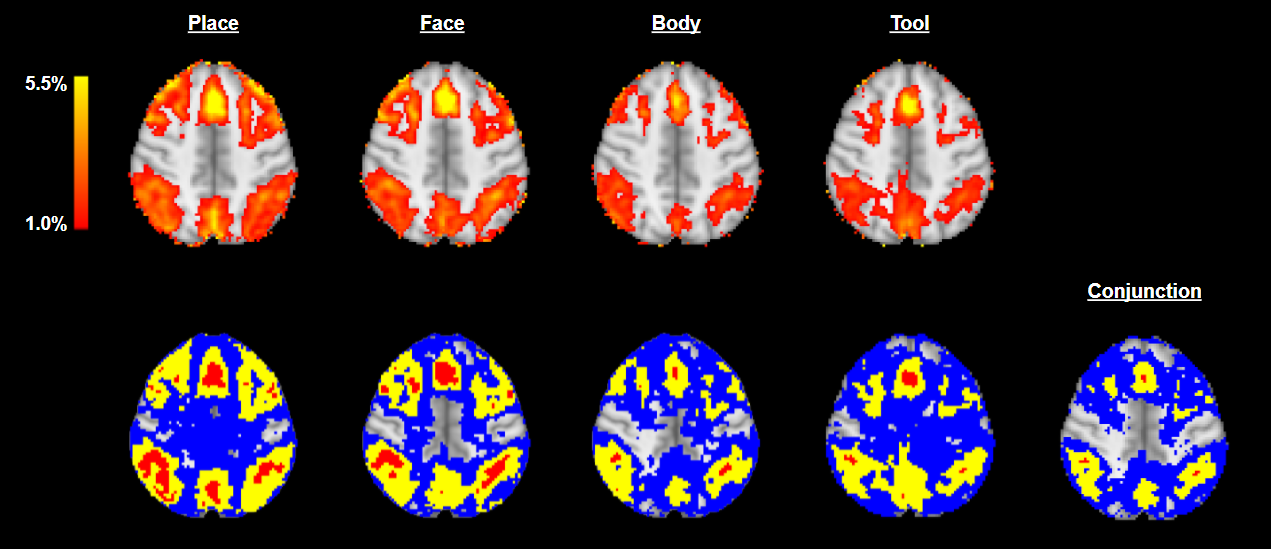}}
\caption{\label{resultsfig} 
$\%$BOLD images (top) and 95\% confidence regions (bottom left) for the HCP working memory task for each of the four visual stimuli types, alongside conjunction confidence regions assessing the overlap of the four thresholded $\%$BOLD images (bottom right). Displayed is axial slice $z=46$mm. The thresholds employed were $c=1\%$ BOLD change, for all images, and $\alpha=0.05$, for the confidence regions. Upper and lower confidence regions are displayed in red and blue, respectively.
Across the bottom row, the yellow sets are the point estimates $\hat{\mathcal{A}}_c^1,...\hat{\mathcal{A}}_c^4$ and $\hat{\mathcal{F}}_c$ respectively. The red conjunction set has localized regions in the Superior Frontal Gyrus (top) and the Angular Gyri (left/right), for which we can assert with $95\%$ confidence that, for all four study conditions, there was (at least) a $1.0\%$ change in BOLD response.}
\end{figure*}

The results are shown in Fig. \ref{resultsfig}. The conjunction inference identified several localized regions, including the Superior Frontal Gyrus, which is well known for its involvement in working memory (c.f. \citet{Boisgueheneuc2006}, \citet{Vogel-2016}, \citet{Alagapan-2019}), and the Angular Gyri, which are well known to be involved in a range of tasks including memory retrieval, attention and spatial cognition (c.f. \citet{Seghier2013}, \citet{Brechet-2018}).
Whilst the confidence regions for conjunction inference illustrated in Fig. \ref{resultsfig} exhibit a clear resemblance to the four contrast images, we note that the red set, $\hat{\mathcal{F}}_c^{+}$, is very small in comparison to the blue set, $\hat{\mathcal{F}}_c^{-}$, which is large and diffuse. This observation conveys important information about the spatial variability of the regions identified. In particular, it can be seen that, in the regions immediately surrounding the Superior Frontal Gyrus and Angular Gyri, there is a much higher resemblance between the blue ($\hat{\mathcal{F}}_c^{-}$) and yellow ($\hat{\mathcal{F}}_c$) sets than is seen across the rest of the brain. This resemblance provides an insight into the degree of spatial variation present surrounding these regions. In this case, the strength and localization of this resemblance suggest that the spatial variability of $\hat{\mathcal{F}}_c$ is less severe in and around the aforementioned anatomical regions than it is across the rest of the brain. It may be concluded that the estimated yellow `blobs' corresponding to the Superior Frontal Gyrus and the Angular Gyri have been more reliably localized than the other yellow `blobs' appearing in the image.

In general, the single-`study condition' confidence regions can be seen to be much larger than those observed for the conjunction inference. This observation matches expectation as the overlap of the four excursion sets is smaller than each set individually. In addition, in this example, the conjunction method employed the entire experimental design for analysis (as opposed to the single-`study condition' method which employed only the data concerning the stimuli of interest) and is therefore expected to have a higher statistical power and, thus, tighter confidence bounds.

\section{Conclusion and Discussion}

In this work, we have produced a method for generating confidence regions for intersections and unions of multiple excursion sets. The confidence regions generated by the method generalise the notion of confidence intervals to arbitrary spatial dimensions and possess the usual frequentist interpretation that, were the $0.95$ procedure repeated on numerous samples, the proportion of confidence regions, $\hat{\mathcal{F}}^{+}_c$ and $\hat{\mathcal{F}}^{-}_c$, that correctly enclose $\mathcal{F}_c$ would tend to $0.95$. Such confidence regions serve as an indicator of the reliability of $\hat{\mathcal{F}}_c$ as an estimate of $\mathcal{F}_c$. Confidence regions which closely resemble the set $\hat{\mathcal{F}}_c$ may be interpreted as signifying that $\hat{\mathcal{F}}_c$ is a reliable estimate of $\mathcal{F}_c$, whilst little resemblance may suggest that there is a high degree of spatial variability present in the data and that the estimated shape, size and locale of $\hat{\mathcal{F}}_c$ are not particularly reliable. Such statements are of particular value in imaging applications where the aim of a statistical analysis is typically to assess how reliably some form of activity may be localized to a particular spatial region.

We stress here that, whilst $\hat{\mathcal{F}}_c$ is equal to the intersection of $\{\hat{\mathcal{A}}_c^i\}_{i \in \mathcal{M}}$, the confidence regions $\hat{\mathcal{F}}^{\pm}_c$ are not the intersections of the single-`study condition' confidence regions $\{\hat{\mathcal{A}}^{\pm,i}_c\}_{i \in \mathcal{M}}$ which would be obtained using the methods of \citet{SSS}. This can be seen by noting that $\{\hat{\mathcal{A}}^{\pm,i}_c\}_{i \in \mathcal{M}}$ are defined using separate thresholds from different bootstraps and can be heavily influenced by the behavior of $\{\partial \mathcal{A}_c^i\}_{i \in \mathcal{M}}$ in spatial regions far from $\mathcal{F}_c$. In fact, treating the naive intersection of the ($1-\alpha$) confidence regions $\{\hat{\mathcal{A}}^{\pm,i}_c\}_{i \in \mathcal{M}}$ as confidence regions for $\mathcal{F}_c$ is not a valid method in general and can result in undesirable asymptotic coverage lying anywhere inside the range $[1-M\alpha,1]$. This claim is further discussed in Supplementary Theory Section S4. To our knowledge, the only valid approach for generating confidence regions for conjunction inference is that outlined in this work.

One potential limitation of the specific methods of Section \ref{LM} is that they allow for confidence regions to be generated only when the target functions, $\{\mu^i\}_{i \in \mathcal{M}}$, are linear combinations of regression parameter estimates. As noted by \citet{Bowring2021}, it is often preferable for an analysis to consider standardized effect sizes, such as Cohen's $d$ (\citet{cohen2013statistical}) or Hedges' $G$ (\citet{hedges}), instead of regression parameter estimates, as standardized effect sizes allow for information about statistical power to be incorporated into the analysis. To this end, \citet{Bowring2021} outlined an approach for generating confidence regions (in the single-`study condition' setting) for images of Cohen's $d$ effect sizes. Here, we suggest that a potential avenue for future work is to combine the methods of \citet{Bowring2021} and those proposed in this work to allow for conjunction, or disjunction, inference to be performed on standardized effect sizes rather than regression parameter estimates. 

Another limitation noted here is that, whilst our proposed method theoretically works for data of arbitrary dimensions, the simulations of Section \ref{sim} verify the performance of our method only for two-dimensional data. Whilst it is typical for many practical applications to focus on two-dimensional data (e.g. climate maps, surface images), higher-dimensional datasets are abundant in a wealth of disciplines (e.g. brain images, astrological maps). For this reason, we intend to investigate further the performance of the method in higher dimensions in future work.

\appendix

\section*{Acknowledgment(s)}

This work was partially supported by the NIH under Grant [R01EB026859] (A.S., T.M., T.N.).

Data were provided by the Human Connectome Project, WU-Minn Consortium (Principal Investigators: David Van Essen and Kamil Ugurbil; 1U54MH091657) funded by the 16 NIH Institutes and Centers that support the NIH Blueprint for Neuroscience Research; and by the McDonnell Center for Systems Neuroscience at Washington University. The code used to obtain the results of Sections \ref{sim} and \ref{RealDat} is freely available and may be found at:\\
\noindent
\textcolor{blue}{\underline{\href{https://Github.com/TomMaullin/ConfSets}{https://github.com/TomMaullin/ConfSets}}}\\
\\
The authors would like to thank Dr. Fabian J.E. Telschow for his early involvement in formulating the project, and Professor Gary R. Lawlor for his email correspondence regarding the differentiability requirements for L'hospital's rule for multivariable functions.

\section*{Disclosure statement}

The authors report there are no competing interests to declare.

\bibliographystyle{plainnat}
\bibliography{Bibliography-MM-MC}
\end{document}